# Theoretical and experimental evidence for a post-perovskite phase of $MgSiO_3$ in Earth's D'' layer.

Artem R. Oganov[1] and Shigeaki Ono[2]

*[1] Laboratory of Crystallography, ETH Zurich, CH-8092 Zurich, Switzerland.*
*[2] Institute for Frontier Research on Earth Evolution, Japan Marine Science & Technology Center, 2-15 Natsushima-cho, Yokosuka-shi, Kanagawa 237-0061, Japan.*

**The Earth's lower mantle, the largest region within our planet (670-2890 km depths), is believed to contain ~75 vol.% of (Mg,Fe)$SiO_3$ perovskite, ~20% (Mg,Fe)O, and ~5% $CaSiO_3$ [1]. This mineralogy cannot explain many unusual properties of the D'' layer, the lowermost ~150 km of the mantle. Here, using *ab initio* simulations and high-pressure experiments we show that at pressures and temperatures of the D'' layer, $MgSiO_3$ transforms from perovskite into a layered $CaIrO_3$–type structure (space group *Cmcm*). The elastic properties of the new phase and its stability field explain most of the previously puzzling properties of the D'' layer: its seismic anisotropy[2], strongly undulating shear-wave discontinuity at its top[3-6], and possibly the anticorrelation between shear and bulk sound velocities[7,8]. This new phase is therefore likely to be a major Earth-forming mineral, and its discovery will change our understanding of the deep Earth's interior.**

If $MgSiO_3$ perovskite is stable throughout the lower mantle, it should be the most abundant mineral in our planet. While some researchers[9] have suggested its decomposition into the oxides at lower mantle conditions, most workers[10-13] found that perovskite is more stable than the oxides. To our knowledge, the possibility that $MgSiO_3$ could be stable in a completely new structure within the lower mantle, has never been considered.

The shear-wave discontinuity at the top of the D'' layer, suggested by Lay and Helmberger[3], has a strong topography. It has been common to explain this discontinuity by some chemical difference between the D'' layer and the rest of the lower mantle. However, using a combination of dynamical and seismic modelling, Sidorin *et al.*[4-6] have shown that the most consistent explanation is a phase transition in mantle minerals. Their best model[5,6] had a shear-wave discontinuity of ~1% located ~150 km above the core-mantle boundary (depth 2740 km), with the Clapeyron slope of 6 MPa/K (though values as large as up to 10 MPa/K were almost equally acceptable). The discontinuities of the compressional wave velocities and of the density could not be resolved. Models of Sidorin *et al.*[4-6] were very appealing, but the major problem was that no appropriate phase transition was known at the time. Here we show that $MgSiO_3$ perovskite undergoes a structural phase transition at the conditions corresponding to the top of the D'' layer. The predicted seismic signatures of this transition match seismological inferences of Sidorin *et al.*[4-6].

A key observation made by Ono *et al.*[14] was that $Fe_2O_3$, like $MgSiO_3$, transforms from the corundum (or ilmenite) to the perovskite structure under pressure. As Ono *et al.*[15] further found, above 60 GPa a post-perovskite phase of $Fe_2O_3$ with a $CaIrO_3$ type *Cmcm* structure[15] (Fig. 1) is stable. This has led to the idea that a similar structure could be stable for $MgSiO_3$ at high pressure.

We explored this idea using *ab initio* simulations based on density functional theory within the local density approximation (LDA) and the generalised gradient approximation[16] (GGA). The calculated enthalpy difference (Fig. 2a) indicates that for pure $MgSiO_3$ the $CaIrO_3$-structured phase becomes thermodynamically more stable than perovskite well within the range of lower-mantle pressures: at 83.7 GPa in the LDA or at 98.7 GPa in the GGA. It is well known[22] that usually the LDA underestimates transition pressures, whereas the GGA slightly (by a few GPa) overestimates them.

The Clapeyron slope was calculated using density-functional perturbation theory[21] to be 9.85 MPa/K within the LDA and 9.56 MPa/K within the GGA. These numbers agree with a simple formula for the high-temperature entropy change[23] for transitions without coordination number changes (as in this case): $\Delta S = 3nk_B\gamma\frac{\Delta V}{V}$, where $n$ is the number of atoms in volume $V$, $\Delta V$ the volume change at transition, $k_B$ the Boltzmann constant, and $\gamma$ the Grüneisen parameter at the transition (1.2 for perovskite[24,25]). This formula yields the Clapeyron slope of ~8 MPa/K.

Following this prediction, we found the new phase experimentally. The sample (pure $MgSiO_3$) was heated with a laser to overcome potential kinetic effects on possible phase transitions. Experimental details have been described elsewhere[13,26]. The experimental powder diffraction pattern is shown in Fig. 3; we could index reflections not belonging to platinum, platinum carbide, and rhenium gasket in the *Cmcm* space group. This is indeed very similar to the theoretical structure (Table 1), thus confirming the predicted stability of the post-perovskite phase in pure $MgSiO_3$. Note that the calculated and experimentally determined stability fields are in excellent agreement with each other (Fig. 2b). At temperatures of the D'' layer (~3000 K) the GGA transition pressure is 127 GPa, which corresponds to the top of the D'' layer, 2740 km depth. We recall that Sidorin *et al.*[4-6] suggested a transition at 127 GPa with a Clapeyron slope of 6 MPa/K.

Equation of state parameters for perovskite and post-perovskite are listed in Table 2, and the elastic constants calculated from stress-strain relations at 120 GPa are given in Table 3. The density discontinuity at the transition is 1.4%; for the mantle the expected density discontinuity is 1.1%. The predicted shear wave discontinuity in $MgSiO_3$ is 1.9% (1.4% for the mantle), consistent with ~1% suggested by Sidorin *et al.*[4-6]. We predict a very small discontinuity for the compressional wave velocity (0.3% for pure $MgSiO_3$), explaining why it was so seldom found at the top of the D'' layer[4-6].

The calculated properties of our new phase explain many mysteries of the D'' layer. In seismological models, horizontally polarized shear waves are faster (by 1% on average[2]) than the vertically polarised ones ($v_{SH}>v_{SV}$). Significant seismic anisotropy of the D'' layer, containing a signature of the convective flow[2], could not be explained even qualitatively by mineral physics[27,28]. The structure of the post-perovskite phase has silicate layers parallel to (010), which is then the most natural slip plane that will be oriented parallel to the convective flow. In regions of horizontal flow (most of D'') using the method of Montagner and Nataf[29] and data of Table 3 for post-perovskite we obtain $\frac{v_{SH}}{v_{SV}} = 1.029 > 1$. This is consistent with seismological evidence, and suggests either a large degree of lattice-preferred orientation or significant contribution from other sources such as shape-preferred orientation (anisotropy due to the ordered distribution of crystals and inclusions in the rock). In regions of upwelling convective streams (e.g., below central Pacific) slip planes would be predominantly vertical and one would obtain $v_{SH}<v_{SV}$, actually observed in such regions[2].

Another mystery of D'', the anticorrelation between the shear ($v_S$) and bulk sound ($v_\phi$) velocities[7,8], can also be quantitatively explained by the phase transition of $MgSiO_3$ from perovskite into the post-perovskite phase. We recall that it has been difficult to explain why at a given depth in the lowermost mantle anomalies of $v_S$ and $v_\phi$ have opposite signs. As Table 3 shows, the post-perovskite transition has a positive jump of $v_S$ and a negative jump of $v_\phi$. Note that the transition is first-order, and in a multicoponent system such as the Earth' mantle there will be an interval $\Delta T$ of coexistence of the two phases at given pressure. In this two-phase region, we approximately write:

$$(\frac{\partial v_\phi}{\partial v_S})_P \approx \frac{(\frac{\partial v_\phi}{\partial T})_{P,x} + f\frac{v_{\phi 2} - v_{\phi 1}}{\Delta T}}{(\frac{\partial v_S}{\partial T})_{P,x} + f\frac{v_{S2} - v_{S1}}{\Delta T}} \quad , \tag{1}$$

which includes purely thermal responses of the velocities and effects due to a phase transition; $f$ is the volume fraction of $MgSiO_3$ (75%). The poorly known effects of variation of Al and Fe content are not included in (1), but we find that anticorrelation can be explained without them. Using formula (1), data of Table 3, and assuming thermal responses equal to those of perovskite[24], we get $(\frac{\partial \ln v_s}{\partial \ln v_\phi})_P$ = -0.15 and -0.33 for $\Delta T$=250 K and 50 K, respectively (from seismic tomography: -0.15[8] and -0.3[7]). Furthermore, one reproduces the positive correlation between the shear and compressional ($v_P$) velocities with the same $\Delta T$: $(\frac{\partial \ln v_s}{\partial \ln v_P})_P$ = 3.36 for $\Delta T$=250 K (3.3 from Ref. 8).

The very fact that so many previously unexplained seismic features of the D'' layer (seismic discontinuity, its magnitude and Clapeyron slope, anisotropy, bulk-shear velocity anticorrelation) are naturally explained by the properties of post-perovskite are a strong indication that this phase is indeed the major component of the D'' layer. The D'' layer is not necessarily chemically very different from the rest of the lower mantle, but it surely is different mineralogically. Neither theoretical and experimental error bars (few GPa for transition pressure) nor the effects of temperature as explored here would change our prediction that the $CaIrO_3$-type post-perovskite phase is stable in the D'' layer. At present very little is known about the effects of $Fe^{2+}$, $Fe^{3+}$, and $Al^{3+}$ impurities on mantle minerals. We expect that since at high pressure $Fe_2O_3$ also has the $CaIrO_3$ structure[14], at least $Fe^{3+}$ impurities should stabilise the post-perovskite phase against perovskite.

One can expect further implications of our findings. For instance, rheological properties of post-perovskite (probably very different from those of perovskite) and the predicted density discontinuity (1.1%) at top of the D'' layer could be important for mantle dynamics. Element partitioning between post-perovskite, perovskite, and molten Fe might be a key to some geochemical anomalies. Elements that are incompatible in the mantle (e.g., Na, K, U, Th) might be easier accommodated in the layered post-perovskite structure, which may affect the chemistry of plume magmas. As $Fe_2O_3$ at high pressure has the same structure as post-perovskite, post-perovskite could have a greater concentration of $Fe^{3+}$ than perovskite. Since the Clapeyron slope of the post-perovskite transition is large, 8.0-9.6 MPa/K, it is likely that the size of the D'' region increased significantly with time as the mantle cooled down. If the whole mantle were molten with temperatures above 4000-4500 K in early history of the Earth, it would be perovskite that crystallised from the cooling melt, and only on further cooling did post-perovskite and the D'' layer appear. The present-day thickness of the D'' could be used to estimate its age, given the cooling history of the lowermost mantle. As post-perovskite stability requires pressures unattainable in smaller planets like Mercury and Mars, many features of these planets would be different from those of the Earth. Further studies are necessary to address these and other (elasticity and anelasticity, electrical conductivity, radiative conductivity, energetics of stacking faults, effects of impurities on stability and properties of post-perovskite) issues.

The results described here break many of the old assumptions and were, in fact, surprising even to us. The post-perovskite phase of $MgSiO_3$ is likely to make a major difference in our understanding of the deep interior of our planet. A lesson that we have learnt through this example is: “If you study the deep Earth, be prepared for surprises”.

**Acknowledgements.** Calculations were performed at CSCS (Manno) and ETH Zurich. We thank P. Ulmer, A.N. Halliday, S. Goes, F. Cammmarano, and A.B. Thompson for fruitful discussions, and Y. Ohishi and N. Sata for experimental support. Synchrotron radiation experiments were performed at the BL10XU, SPring-8. During the refereeing process, an independent experimental study was published [30]. Its results are consistent with our theoretical and experimental findings of the post-perovskite phase.

---

Correspondence and requests for materials should be addressed to A.R. Oganov (a.oganov@mat.ethz.ch)

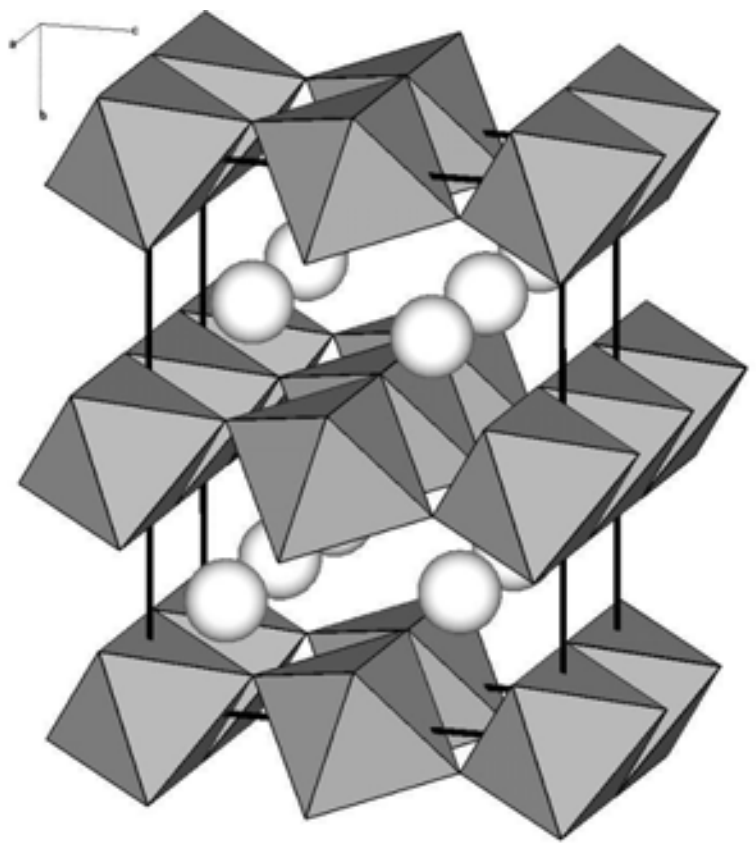

**Fig. 1- Oganov & Ono**

FIG. 1. Structure of post-perovskite phase of $MgSiO_3$ (calculated at 120 GPa). $SiO_6$ octahedra and Mg atoms (spheres) are shown. Similar structures are known for $Fe_2O_3$, $CaIrO_3$, $FeUS_3$, $PbTlI_3$, $UScS_3$, $KTmI_3$, $AgTaS_3$, $CaInBr_3$.

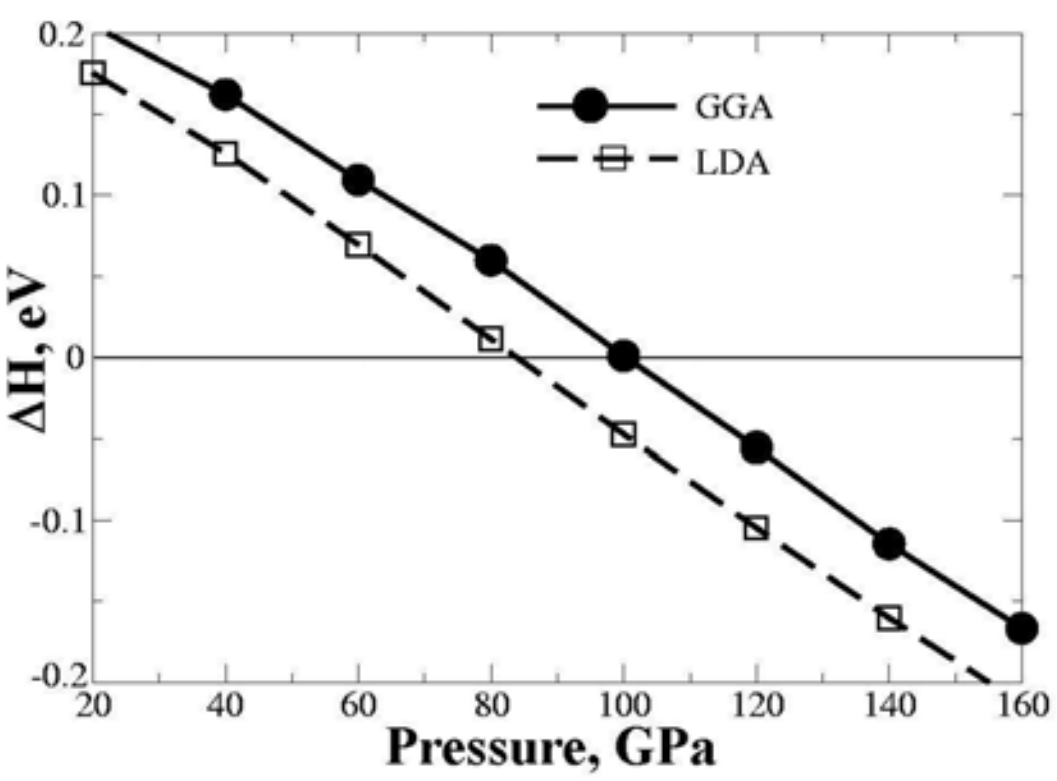

**Fig. 2a - Oganov & Ono**

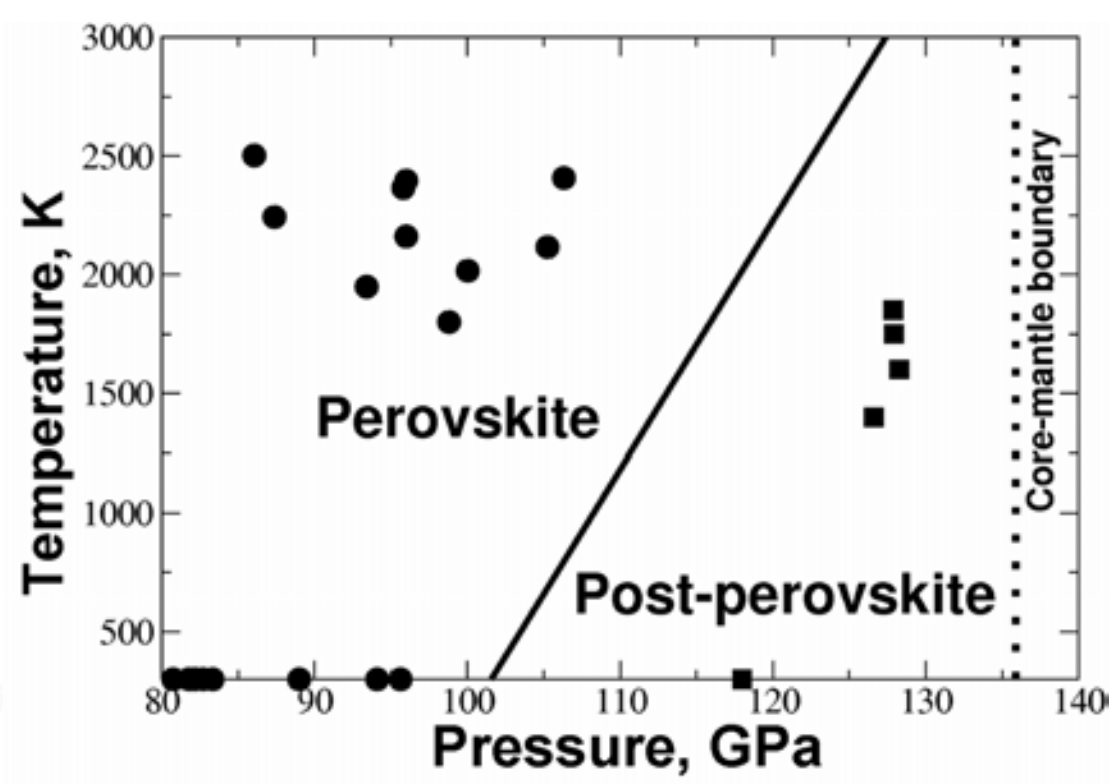

**Fig. 2b - Oganov & Ono**

FIG. 2. Stability of the post-perovskite phase. (a) Static enthalpy difference showing a phase transition at 98.7(83.7) GPa in the GGA(LDA). GGA calculations used the all-electron PAW method[17,18] as implemented in the VASP code[19], with $1s^2 2s^2$ core (radius 2 a.u.) for Mg, $1s^2 2s^2 2p^6$ core (radius 1.5 a.u.) for Si, $1s^2$ core (radius 1.52 a.u.) for O. With the plane wave cut-off of 500 eV and the Brillouin zone sampled by Monkhorst-Pack grids, 4x4x4 for perovskite and 6x6x4 for post-perovskite (for the latter we used the primitive cell with 10 atoms), total energy differences and pressure are converged to 0.006 meV/atom and 0.25 GPa, respectively. Structural relaxation was done using conjugate gradients, until the total energy changes were below $10^{-4}$ eV. The fully converged LDA calculations were done using the ABINIT code[20] (ABINIT is a common project of the Université Catholique de Louvain, Corning Inc., and other contributors, URL http://www.abinit.org). These calculations used Troullier-Martins pseudopotentials, plane waves with the cut-off of 1088 eV, and the same Monkhorst-Pack grids as in the GGA case.

(b) *p-T* phase diagram: theory and experiment. Experimental points: perovskite (Ref. 13) for mantle composition (KLB-1 peridotite), post-perovskite for pure $MgSiO_3$. The solid line is based on the static transition pressure (98.7 GPa) and the Clapeyron slope of 9.56 MPa/K calculated using density-functional perturbation theory and the GGA, with the ABINIT code[20]. In these calculations, the dynamical matrices were calculated on 2x2x2 and 3x3x2 grids in the Brillouin zone for perovskite and post-perovskite, respectively, using density-functional perturbation theory[21]. Interpolating these throughout the Brillouin zone, we calculated phonon frequencies at a very dense reciprocal-space mesh. Errors in the calculated frequencies are within 5 $cm^{-1}$ (mostly less than 1 $cm^{-1}$). From the resulting phonon spectra we calculated various thermodynamic properties, including the entropy (S) of each phase and the high-temperature Clapeyron slope $\frac{dP}{dT} = \frac{\Delta S}{\Delta V}$ .

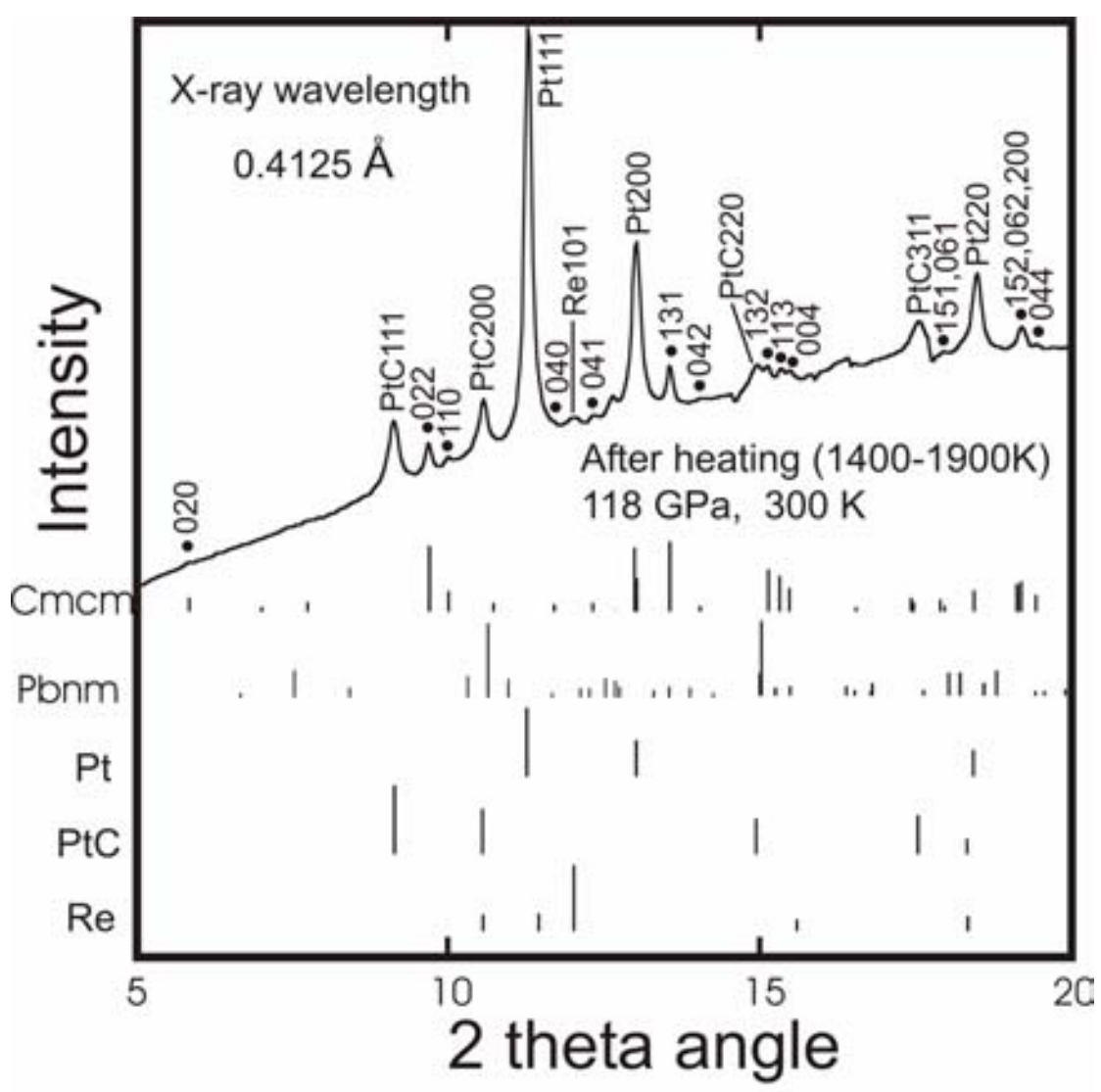

Fig. 3 - Oganov & Ono

FIG. 3. Experimental powder diffraction pattern at 118 GPa and 300 K matching the predicted lattice parameters. The sample was heated with a TEM01-mode YLF laser from two side without the pressure transmitting medium. The pressure was determined using the reference equation of state of platinum (which was mixed with the sample). The sample temperatures were measured from two sides using the spectroradiometric method. The samples were probed using an angle-dispersive X-ray diffraction technique at the synchrotron beam lines BL10XU, SPring-8 in Japan. Solid circles: post-perovskite; Pt: platinum used as an internal pressure calibrant; Re: rhenium of the gasket; PtC: platinum carbide. Observed unit cell dimensions of post-perovskite are a = 2.471(4) Å, b =8.091(12) Å, c = 6.110(11) Å, V = 122.16(34) $Å^3$. Observed volume change form perovskite to post-perovskite is 0.9 %. Vertical bars indicate the calculated positions of the diffraction lines of *Cmcm* post-perovskite phase, *Pbnm* perovskite, platinum, NaCl-type platinum carbide, and rhenium gasket.

**Table 1. Structures of post-perovskite and perovskite at 120 GPa (GGA results).**

| Post-perovskite. Space group *Cmcm*, *a*=2.474, *b*=8.121, *c*=6.138 Å. | | | | Perovskite. Space group *Pbnm*, *a*=4.318, *b*=4.595, *c*=6.305 Å. | | | |
|---|---|---|---|---|---|---|---|
| **Mg** | 0 | 0.2532 | ¼ | **Mg** | 0.5246 | 0.5768 | ¼ |
| **Si** | 0 | 0 | 0 | **Si** | ½ | 0 | ½ |
| **O1** | 0 | 0.9276 | ¼ | **O1** | 0.1164 | 0.4669 | ¼ |
| **O2** | 0 | 0.6356 | 0.4413 | **O2** | 0.1829 | 0.1926 | 0.5575 |
| Distances (in Å): Mg-O1=1.880 (x2), Mg-O2=1.955(x4), 2.099(x2); Si-O1=1.643 (x2), Si-O2=1.695 (x4). | | | | Distances (in Å): Mg-O1=1.833(x1), 1.893(x1), Mg-O2=1.864(x2), 2.047(x2), 2.201(x2); Si-O1=1.661(x2), Si-O2=1.659(x2), 1.670(x2). | | | |

**Table 2. Vinet equations of state of perovskite and post-perovskite.**

| Parameters | **Post-perovskite** | | Perovskite | | |
|---|---|---|---|---|---|
| | LDA | GGA | LDA | GGA | Exp.* |
| $V_0$, Å$^3$ | 162.86 | 167.64 | 163.35 | 167.42 | 162.3 |
| $K_0$, GPa | 231.93 | 199.96 | 259.82 | 230.05 | 259.5 |
| $K_0^{'}$ | 4.430 | 4.541 | 4.060 | 4.142 | 3.69 |

* Third-order Birch-Murnaghan equation of state parameters at room temperature[10].

**Table 3. Elastic constants of perovskite and post-perovskite at 120 GPa*.**

| | $C_{11}$ | $C_{22}$ | $C_{33}$ | $C_{12}$ | $C_{13}$ | $C_{23}$ | $C_{44}$ | $C_{55}$ | $C_{66}$ | $K$ | $G$ |
|---|---|---|---|---|---|---|---|---|---|---|---|
| **Perovskite** | 907 | 1157 | 1104 | 513 | 406 | 431 | 364 | 271 | 333 | 648.0 | 310.9 |
| | Acoustic velocities: $v_p$=14118, $v_s$=7636, $v_\phi$=11026 m/s | | | | | | | | | | |
| **Post-perovskite** | 1252 | 929 | 1233 | 414 | 325 | 478 | 277 | 266 | 408 | 647.2 | 327.5 |
| | Acoustic velocities: $v_p$=14158, $v_s$=7783, $v_\phi$=10940 m/s | | | | | | | | | | |

*GGA results. All elastic constants are in GPa.